\documentclass{article}
\usepackage{ijcai24}

\usepackage{times}
\usepackage{soul}
\usepackage{url}
\usepackage[hidelinks]{hyperref}
\usepackage[utf8]{inputenc}
\usepackage[small]{caption}
\usepackage{graphicx}
\usepackage{amsmath}
\usepackage{amsthm}
\usepackage{booktabs}
\usepackage{algorithm}
\usepackage{algorithmic}
\usepackage[switch]{lineno}
\usepackage{comment}
\usepackage{enumitem}
\usepackage{wrapfig}
\title{Agent-based Modeling meets the Capability Approach for Human Development: \\ Simulating Homelessness Policy-making}

\author{
Alba Aguilera $^1$
\and
Nardine Osman$^1$\And
Georgina Curto$^{2}$
\affiliations
$^1$Artificial Intelligence Research Institute, IIIA-CSIC, Barcelona, Spain\\
$^2$United Nations University Institute in Macau, Macau SAR, China\\
\emails
\{aaguilera, nardine\}@iia.csic.es,
curto@unu.edu
}

\begin{document}

\maketitle
%While most capability theories have been used for normative and evaluative purposes, this paper adopts an explanatory approach
\begin{abstract}
%The global rise in homelessness calls for urgent and alternative policy solutions. Non-profits and governmental organizations alert about the many challenges faced by people experiencing homelessness (PEH), which include not only the lack of shelter but also severe forms of discrimination. In this context, the capability approach (CA), which underpins the United Nations Sustainable Development Goals (SDGs), provides a comprehensive framework to assess inequity in terms of resources,  human agency and opportunities. This paper explores how agent-based modelling and reinforcement learning can be combined with the CA to: (1) define agents' behavior towards restoring their central capabilities, considering human values and contextualized needs, and (2) evaluate specific social  policies by generating agent-based simulations. The framework is developed in a real case study of health inequity and homelessness, working in collaboration with stakeholders, nonprofits and domain experts. The ultimate goal of the project is to present a novel CA-based agent model framework that can be replicated in a diversity of social challenges to evaluate alternative inequity mitigation policies in a non-invasive way.  
The global rise in homelessness calls for urgent and alternative policy solutions. Non-profits and governmental organizations alert about the many challenges faced by people experiencing homelessness (PEH), which include not only the lack of shelter but also the lack of opportunities for personal development. In this context, the capability approach (CA), which underpins the United Nations Sustainable Development Goals (SDGs), provides a comprehensive framework to assess inequity in terms of real opportunities. This paper explores how the CA can be combined with agent-based modelling and reinforcement learning. The goals are: (1) implementing the CA as a Markov Decision Process (MDP), (2) building on such MDP to develop a rich decision-making model that accounts for more complex motivators of behaviour, such as values and needs, and (3) developing an agent-based simulation framework that allows to assess alternative policies aiming to expand or restore people's capabilities. The framework is developed in a real case study of health inequity and homelessness, working in collaboration with stakeholders, non-profits and domain experts. The ultimate goal of the project is to develop a novel agent-based simulation framework, rooted in the CA, which can be replicated in a diversity of social contexts to assess policies in a non-invasive way. %Our ultimate goal is to help identify the optimal policies that expand basic capabilities and functionings. %prioritised in the context of homelessness. .including multidimensional health issues and significant barriers to accessing healthcare services..

\end{abstract}

\section{Introduction}
%\textbf{Problem Statement.}
According to a recent report by the Organization for Economic Cooperation and Development (OECD), approximately 2.2 million people experience homelessness in its 35 member countries %and this is an underreported and growing social challenge
~\cite{oecd2024homelessness}. In the European Union, it is estimated that on any given night, at least 895,000 people sleep on the streets~\cite{feantsa2023overview}. In Barcelona, Spain, the number of people sleeping rough increased from 658 in 2008 to 1,063 in 2022~\cite{diagnosi2022barcelona}. Similar trends are observed in London and San Francisco, where the number of people experiencing homelessness (PEH) has tripled and quadrupled~\cite{trustforlondon2023,sf_homeless_population_2024}. 

The increasing scale of homelessness demands for urgent and alternative policy-making. European institutions have called for a shift from traditional homelessness management to a model that aims to actually solve this complex social challenge by offering comprehensive support to PEH~\cite{EuropeanCommission_Homelessness}. This aligns with the Universal Declaration of Human Rights~\cite{un1948udhr} and the Capability Approach (CA), which underpins the United Nations Sustainable Development Goals (SDGs)~\cite{gasper2017sdgs}. Unlike traditional development models, which focus on the amount of resources available to individuals (e.g. utilitarian frameworks), the CA shifts the attention to the people's ability to transform these resources into opportunities or capabilities (what people are actually able to do and be). From this perspective, homelessness and poverty can be understood as severe forms of capability deprivation~\cite{marshall2024beyond}, where individuals are unable to access their central capabilities (in Table~\ref{tab:nussbaum_capabilities}). Central capabilities are a list of ten 'essential human entitlements to be able to conduct a meaningful life with dignity'~\cite{nussbaum2011}, such as being able to be adequately nourished or sheltered, being secured against violent assault, or being able to undertake employment. Providing the required support to PEH requires not only the provision of material resources (such as food or shelter), but also social and institutional arrangements aiming to restore and expand these central capabilities~\cite{sen1999}. %that facilitate the conversion of resources into opportunities,

\begin{table*}[ht!]
\centering
\renewcommand{\arraystretch}{1.2}  % Increase row spacing for readability
\begin{tabular}{p{3cm} p{14.5cm}}
\hline
\textbf{Life} & 
Being able to live to the end of one’s lifespan without premature death \\
\textbf{Bodily health} & 
Being in good physical health, including reproductive health \\
\textbf{Bodily integrity} & 
Being able to move freely, being free from violence, having bodily, reproductive, and sexual autonomy \\
\textbf{Senses, imagination, thought} & 
Being able to reason, think, and create; having access to art, literature, and science; and enjoying pleasurable experiences while avoiding non-beneficial pain \\
\textbf{Emotions} & 
Being able to form and mourn emotional attachments to others \\
\textbf{Practical reason} & 
Being able to conceptualize what is good and plan one’s future \\
\textbf{Affiliation} & 
Subdivided into interactions with others and dignified, non-discriminatory participation in society \\
\textbf{Other species} & 
Being able to live with concern for animals, plants, and the natural world \\
\textbf{Play} & 
Laughter, play, and recreational activities \\
\textbf{Control over one’s environment} & 
Subdivided into political participation and material rights to own property and undertake employment \\
\bottomrule
\end{tabular}
\caption{Central Capabilities, adapted from~\protect\cite{nussbaum2011}.}
\label{tab:nussbaum_capabilities}
\end{table*}
Non-profit and governmental organizations are responding to this social challenge with transformative policy proposals. For instance, in Barcelona, the “Proposed Law on Transitional and Urgent Measures to Address Homelessness” ~\cite{sjd2023} is currently under discussion. So we are collaborating with domain experts and non-profit organizations (namely Arrels Fundació, Caritas and Salut Sense Llar), as well as with human development academics specialized in poverty (namely from the The Sustainability, Economics and Ethics (SEE) group). Our research aims to provide empirical results that guide the evaluation and implementation of these novel policy approaches. 

To achieve this, we propose building a novel social simulation tool rooted in the CA. This is an innovative and interdisciplinary project, which draws on expertise from human development and computer science to explore the application of the CA in the agent-based modelling (ABM) and reinforcement learning domain. Additionally, we integrate complex motivators of behaviour, such as human values~\cite{schwartz} and needs, into the CA. This seeks to enhance the modelling of human behavior in the context of homelessness and contribute to prior research into the computational representation of human values~\cite{nardine}. % that contemplate the multidimensions of homelessness

The goals of the project are as follows: 

\begin{enumerate}[label=\arabic*)]
    \item Implement the framework of the CA as a Markov Decision Process (MDP) that can be applied across social contexts where inequity arises, such as homelessness.
    \item Building on such MDP, develop a novel and rich agent decision-making model that accounts for complex motivators of behaviour, including human values and needs. 
    \item Develop an agent-based simulation framework to assess the impact of policies on people's capabilities and identify those that effectively restore or expand them, working towards the UN SDGs.
\end{enumerate}

Our project will leverage agent-based modelling (ABM), which in the context of homelessness, allows to simulate the behavior of a diversity of stakeholders involved (including PEH, social workers and non-profits), interacting with each other autonomously. ABM will facilitate (1) the exploration of outcomes through both top-down (impact of legal policies) and bottom-up (impact of resources or disparities) processes, (2) the modelling of individual behaviour, social interactions and the role of policies in restoring their central capabilities and (3) the evaluation of both of the above from the perspective of equity~\cite{williams2022integrating}. In this paper, we define equity and inequity in terms of having (or lacking) basic capabilities~\cite{sen1979} to conduct a meaningful life with dignity~\cite{sen1999}, in line with the CA. We acknowledge complementary interpretations of inequity, on the basis of distribution of income and wealth~\cite{piketty2017}, oppression and social justice~\cite{sen1979}, the equality of opportunities and meritocracy~\cite{sandel2020}, social and political recognition~\cite{honneth1996,taylor1989}, prejudices and marginalization of minority groups~\cite{allport1954}, or the representation in political and economic institutions~\cite{acemoglu2012}, among others. %Moreover, compared to other modelling approaches, ABM offers unique advantages to incorporate equity considerations~\cite{williams2022integrating}. We can represent the disparities within a population, the agency of individuals, and the resulting behaviour emerging from their interactions. Therefore, ABM allows us (1) to investigate the emergence of outcomes through both top-down (impact of legal policies) and bottom-up (impact of resources) processes, (2) to model individual behaviour, affected by values, needs and capabilities, where policies help them restore their central capabilities and (3) to evaluate both of the above from the perspective of equity. In this paper, we define equity and inequity in terms of having (or lacking) basic capabilities~\cite{sen1979} to conduct a meaningful life with dignity~\cite{sen1999}. We acknowledge complementary interpretations of inequity, on the basis of distribution of income and wealth~\cite{piketty2017}, oppression and social justice~\cite{sen1979}, the equality of opportunities and meritocracy~\cite{sandel2020}, social and political recognition~\cite{honneth1996,taylor1989}, prejudices and marginalization of minority groups~\cite{allport1954}, or the representation in political and economic institutions~\cite{acemoglu2012}, among others.  

\section{Related Work}

\subsection{The Capability Approach}
\begin{figure*}[ht]
    \centering
    \includegraphics[width=0.8\linewidth]{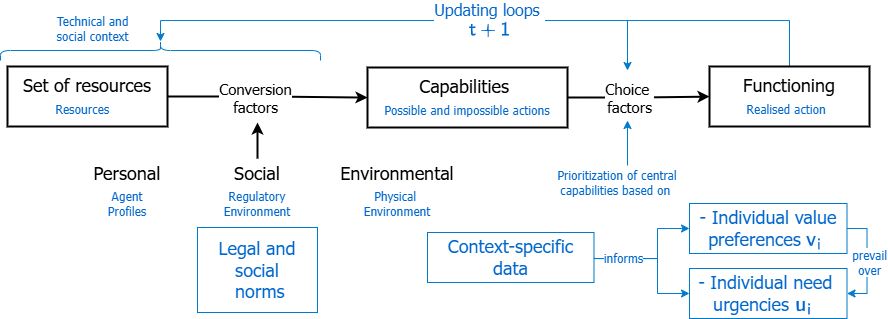}
    \caption{Schematic overview of the Capability Approach~\protect\cite{robyens2017} (in black), with the integration of computational elements to operationalize it in the agent-based modelling domain (in blue).} %Among them, we highlight the integration of legal and social norms~\cite{Shoham1995}, as well as collective contextualized data informing individual needs and values~\cite{schwartz}. } %, to operationalize it in the agent based modelling domain.
    \label{fig:CAvalues}
\end{figure*}

The capability approach (CA), originally developed by Amartya Sen~\cite{sen1999} and further expanded by Martha Nussbaum~\cite{nussbaum2011}, is a flexible and multi-purpose framework for
analyzing how to promote and assess
human well-being, development, and social
justice. Rather than looking only at resources or outcomes (functionings), it focuses on the opportunities (capabilities) people have to lead the lives they value. 
Following~\cite{robyens2017} terminology we define the key terms for computational purposes, illustrated in black in Fig.~\ref{fig:CAvalues}:

\begin{enumerate}
    \item \textbf{Resources}: set of commodities and services that are available to a person in a given context, such as a bike, adequate shelter, or public healthcare.  %They are only useful when paired with appropriate conversion factors.  of rights, entitlements and

    \item \textbf{Conversion Factors}: Characteristics that facilitate the transformation of resources into capabilities. These include personal (such as physical condition, reading skills or gender), social (such as social norms, public policies or discrimination) and environmental (such as climate, infrastructures or transportation).
    % \begin{enumerate}
    % \item Personal factors internal to the person, such as metabolism, physical condition, sex, reading skills, or intelligence. 
    % \item Social factors from the society in which one lives, such as \emph{social norms}, \emph{public policies}, discrimination, or societal hierarchies. 
    % \item Environmental factors from the physical and built environment in which a person lives, including climate, pollution, infraestructure, or transportation.
    
%\end{enumerate}
In our work, social conversion factors will be interpreted as social and legal norms in line with the social laws of~\cite{Shoham1995}. We highlight that these norms could also serve as proxies for institutional discrimination~\cite{aguilera2024can}.

\item \textbf{Capabilities}: what people are able to do and be given their resources and conversion factors. In our work, it will be useful to distinguish between the following:
    \begin{enumerate}
        %\item \textbf{Basic vs Central}: Basic capabilities refer to the freedoms to do some basic things considered necessary for survival and to avoid or escape poverty or other serious deprivations. Martha Nussbaum defined a list of ten basic capabilities, denominated central capabilities, that everyone should be entitled to, as a matter of human dignity. 
        \item \textbf{Central capabilities:} Ten basic capabilities that everyone should be entitled to as a matter of human dignity~\cite{nussbaum2011}, defined in Table~\ref{tab:nussbaum_capabilities}.  
        \item \textbf{Specific capabilities:} Context-dependent capabilities. For instance, ‘being able to move from a peripheral neighbourhood to work’ can be considered the more specific capability for ‘being able to move freely from place to place’, related to the central capability of \textit{bodily integrity} in Table~\ref{tab:nussbaum_capabilities}.
        %The ends of policy making and institutional design is to provide people with general capabilities, whereas the ends of persons are more specific capabilities. 
    \end{enumerate}
In line with~\cite{robyens2017}, central capabilities will be the ends of policy-making and part of the evaluative space for assessing inequity in our work. Specific capabilities will be the contextual ends or goals of individuals, represented as possible actions, and often serving as means for enabling central capabilities. This follows the means-ends distinction in the CA, which differentiates between capabilities as ends or as means to achieve other ends. The idea is that by acting upon specific capabilities (realising actions), individuals can restore central ones, because capabilities are interconnected: achieving one can enable others (both for an individual's own development or a society's development). %This is represented in the dynamics of our decision-making as an updating loop between realised actions and resources or conversion factors.
\item \textbf{Choice Factors}: Motivators of behaviour influencing the indidivual's choice of acting upon specific capabilities (preferring to perform one action over another). %, such as values, beliefs, needs or personality traits.

\begin{quote}
\small

\textit{“There is very little about these constraints that one could say in general terms, as they are so closely interwoven with a person’s own history and thus with her personality, emotions, values, desires and preferences.”}~\cite{robeyns2005capability}
\end{quote}
    Given the underspecified role of choice factors in the CA literature, our work incorporates individual needs and values into the framework, drawing on context-specific sources and established theories, such as Schwartz’s value framework~\cite{schwartz}.%As noted in~\cite{robyens2017}, "there is very little about these constraints that one could say in general terms, as they are so closely interwoven with a person’s own history and thus with her personality, emotions, values, desires and preferences.”  %These are not explicitly tackled in the core literature of CA, but in Schwartz's framework, .... value promotion...
    \item \textbf{Functionings}: Realised actions, what people end up doing and being in practice. %the set of beings and doings that a person ultimately puts into practice. %These include working, resting, being literate, being healthy, or being part of a community.
 
\end{enumerate}

In the context of inequity, poverty and homelessness, it makes sense to define behaviour not only in terms of \emph{what people value or need to do and be}, but also in terms of \emph{what people can actually do and be}. The capability approach addresses these fundamental questions: what are people actually able to do and be, given their circumstances? And how individual capabilities can be expanded so that all human beings can conduct a meaningful life with dignity? It emphasizes that our action space is constrained by resources and conversion factors, i.e. there are things we cannot do because we lack the means or the personal, social and environmental circumstances to do them. Capabilities provide this counterfactual information that is normally overlooked in computational decision-making architectures: they represent the set of an individual's impossible and possible actions to be realised. The choice of which possible action we realise in a particular instant of time (which capability is transformed into a functioning), depends on individual goals and motivators of behaviour (personal choice factors).%the personal choice factors.  %should we say goals here???
%In line with other mathematical notation in the literature~\cite{chavez2021capmod}, we consider the set of resources available to an individual $i$ as $x_i$. Conversion factors are the characteristics determining how resources are converted into vectors of potential functionings, denoted as $q_i$, within a capability set. Meanwhile, choice factors act as prioritisation mechanisms, influencing how capabilities are prioritised to be converted into a vector of functionings, denoted as $b_i$, at a certain instant of time. %$q_i \in Q_i$
%capabilities are what people are able to do given their resources and circumstances.  %This whole scheme is illustrated in black in Fig.~\ref{fig:CAvalues}. Unlike other development models that focus on the amount of material wealth and resources available to individuals (i.e. traditional utilitarian frameworks), the CA focuses on how these resources can be converted to pursue meaningful functionings (states of being or doing).
%Conversion factors are a vector of XXX that apply to an individual $i$ , $c_i$. 
%A conversion FUNCTION is then a function $q : R \times F \times C \to [0,1]$, where R is a set of resources, F is a set of conversion factors and C is a set of capabilities, and $q$  defines the probability the r $\in$ R and $cf$ $\in$ F contribute to achieving c $\in$ C.

\subsection{The CA in Agent-based Modelling}
There is an important body of ABM literature focusing on equality, equity and fairness in different practical domains. However, to the best of our knowledge, few publications in the agent-based modelling literature use the capability approach as a conceptual framework for social simulation. The existing literature that does, can be broadly divided into two research lines: (i) using ABMs as a tool to infer capabilities for CA purposes, and (ii) using the CA as a conceptual framework for designing ABMs. 
 
The first line of research is a very small body of work that combines ABM with Structural Equation Modelling (SEM)~\cite{chavez2021capmod}. Although related, their main purpose and methods are not aligned with our research. The second line of research involves creating ABMs for practical domains, such as in energy justice~\cite{melin2021energy,assa2020can,de2020conflicted} and community resilience~\cite{markhvida2020quantification,silva2022commuter,tseng2024ci}. These studies mainly examine how capabilities and functionings (in the evaluative space) are affected by different resources or conversion factors. While in~\cite{tseng2024ci} the simulation is rooted in the CA (as we intend to do), their choice factor is entirely dependent on~\cite{maslow1943needs}'s hierarchy of needs. This may be suitable for their context, but we consider it oversimplifies the CA's core principle, which underscores that all capabilities are essential and does not prescribe a universal hierarchy of prioritisation. %, as such prioritisation must be adapted to specific circumstances

Our framework aims to address this gap by including more complex motivators of behaviour in the choice factor, such as value and need preferences, informed by context-specific data. This aims to provide a robust decision-making that can be applied across contexts where inequity arises. We emphasize the need for interdisciplinary collaboration to define all relevant aspects of the approach by following~\cite{robyens2017}'s modular view of the CA. We work towards its operationalization in the computational domain, using an MDP as the technical basis of the decision-making. 

\section{Methods}

Before presenting our proposal, we begin with an example that highlights our motivation and objective. From there, we explain: 1) how the CA and its notions, such as conversion factors, capabilities or choice factors (including values and needs) can be implemented computationally, 2) how can we develop a novel and rich agent decision-making model by mapping the CA to an MDP; and 3) how to develop an agent-based simulation framework that allows us to evaluate the impact of policies aiming to expand or restore people's capabilities, in line with the UN SDGs. %: 1) how the CA and its notions, such as resources, conversion factors, capabilities, etc., along with motivators of behaviour like values and needs, can be implemented as an MDP; 2) how to build on such an MDP to develop a novel and rich agent decision-making model that goes beyond traditional ones by accounting for more complex motivators of behaviour; and 3) how to develop an agent-based simulation framework that allows us to evaluate the impact of policies aiming to expand or restore people's capabilities, in line with the UN SDGs.  %which is one of the capability approach's main objectives. For example, the simulation could help policy makers analyse policies and their impact on improving human capabilities.%how capabilities, values and needs can be integrated into a decision-making architecture in the agent-based modelling domain. %Next, we describe the technical foundations of decision-making in reinforcement learning and we conceptually map this with the CA for its application. %  supported by literature from both CA and human value theories.

\subsection{Example: The Difference a Bike Makes} 
Let's consider a simple and practical example in the CA framework: a person with access to a bike. A bike, in general, should provide a person with the ability to move freely and faster than walking. However, the person's ability to ride a bike is not only dictated by available resources, but by several conversion factors that vary from one person and context to another. For instance, (i) if the individual has a disability or doesn’t know how to ride a bicycle (personal factors), (ii) if riding a bicycle is forbidden or discouraged by social or legal norms (social factors), or (iii) if the individual lives in a snowy place that renders riding a bike impossible (environmental factors). Conversion factors help us define the probability of a capability being enabled from available resources. This probability can either be binary, completely conditioning whether a person moves or not with 100\% or 0\% probabilities, or non-binary, partially conditioning a person's mobility. Note that if `riding a bike' is a possible action, choosing to perform that action may act as a mean for enabling other actions, such as `securing a job'. All these actions can be related to central capabilities, such as \textit{bodily integrity} or \textit{control over one's environment} in Table~\ref{tab:nussbaum_capabilities}. In this way, possible actions can serve as evaluation criteria to assess central capabilities. %, considered as the long-term goals behind the actions of the individual. %Both of these specific capabilities (safely riding a bicycle or securing a job), can be related to central ones~\cite{nussbaum2011}. %. We note that if the person chooses to perform that action, realising that action may act as a mean for enabling other actions, such as "securing a job". These unlocked possible actions can be related to central capabilities in Table~\ref{tab:nussbaum_capabilities}, such as control over one's environment", serving as evaluation criteria to assess how capabilities expand. %Both of these specific capabilities (safely riding a bicycle or securing a job), can be related to central ones~\cite{nussbaum2011}. 

Choosing to realise an action is driven by personal choice factors, such as values and needs. For instance, in situation (ii), where riding a bicycle is discouraged by our society, a strong alignment with \textit{tradition} and \textit{conformity} values may lead a person to prefer walking over biking, even if walking takes significantly more time. On the other hand, someone valuing \textit{stimulation} over \textit{tradition} may choose to bike, despite social and legal norms or (iii) hazardous snowy conditions, simply for the thrill and adventure. Our framework aims to build the decision-making taking such personal choice factors into account. %These value preferences can be represented as a vector, conditioning which is the most valued capability at a certain instant of time. %We can model this simple problem as an MDP.can make someone go over the speed limit or forget to wear a helmet  Additionally, the need to arrive somewhere on time can add a layer of urgency in our decision-making. 

%Considering human values as proxy of choice factor, we can describe several situations where different value preferences influence the achieved functioning. For instance, if because of cultural values we face situation (ii), where riding a bicycle is forbidden or discouraged by societal norms, a strong sense of tradition and conformity may lead the individual to prefer walking over biking, even if walking takes significantly more time. On the other hand, someone valuing stimulation may choose to bike, despite societal norms or (iii) hazardous conditions, simply for the thrill or adventure. These value preferences can be represented as a vector, conditioning which is the most valued capability at a certain instant of time. %We can model this simple problem as an MDP.

\subsection{Proposed Choice Factors for an Enhanced Decision-Making Model}

\begin{table*}[ht]
    \centering
    \renewcommand{\arraystretch}{1.0}
    \begin{tabular}{|p{2.5cm}|p{7.5cm}|p{6.5cm}|} \hline
         \textbf{Concept} & \textbf{Capability Approach} & \textbf{Markov Decision Process} \\ \hline
         Input and output & Resources, Personal characteristics, Need and value preferences & States \\ \hline
         From resources to capabilities& Conversion Factors & Transition Probabilities \\ \hline
         From capabilities to functionings
         %, feedback loop between functionings and choice factors
         & Choice Factors (need urgencies and value preferences)& Short and Long-term Rewards  \\ \hline
Behaviour
    & \begin{itemize}[leftmargin=*, itemsep=0.1pt, topsep= 0pt]
        \item Specific capabilities 
        \item Deprived specific capabilities 
        \item Specific functionings
      \end{itemize}
    & \begin{itemize}[leftmargin=*, itemsep=0.1pt, topsep= 0pt]
        \item Possible actions
        \item Impossible actions
        \item Realised actions
      \end{itemize} \\
    \hline
 % Interconnectedness of capabilities or means and ends distinction & Feedback loop between functionings and resources or conversion factors & Feedback loop between actions and states or transition probabilities\\\hline
 Output & Central capabilities and functionings & Long-term goals %(enabling central capabilities) 
 \\\hline
    \end{tabular}
    \caption{Mapping between the Capability Approach and a Markov Decision Process. }
    \label{tab:my_label}
\end{table*}

Different motivators of behaviours have been considered in the literature. Values, according to~\cite{schwartz2012overview}, ``refer to desirable goals that motivate action''. Needs, as defined by~\cite{maslow1943needs}, have also been considered as basic motivators of action. Their connection to values has been previously established. For instance, in~\cite{dignum2021social}, the authors recognize that values inform the prioritisation of needs, but behaviour is primarily driven by Maslow's needs. %universal to all human beings, but varying in importance among individuals and groups
%According to Schwartz's framework~\cite{schwartz}, values are dynamic and contextual, i.e. they evolve over time and change across cultures. Values inform the prioritisation of needs, which are universal to all human beings but vary in importance and urgency among individuals and groups~\cite{dignum2021social}. 

\textbf{Proposed Choice Factors. }In our work, needs give an indication of \emph{urgency} and values give an indication of \emph{importance} to our actions or goals. We differentiate between urgency-based actions (or short-term goals), that we perform for the short-term reward, and importance-based actions (or long-term goals) performed for the long-term reward. For instance, in the context of homelessness, a person might urgently seek shelter (urgency-based action), yet choose to remain disengaged from social services (importance-based action). Although accessing shelter could provide immediate short-term rewards, a strong alignment with certain values, such as \textit{security}, may prevail over the need, prioritizing long-term rewards instead. We treat values as prevailing over needs, although the detailed interactions between them will be explored in our proposed work.

Capturing individuals' value preferences and need urgencies is usually a complex task. One way to address this challenge would be to rely on questionnaires with the PEH to understand their priorities in a given context (e.g. in terms of the central capabilities in Table~\ref{tab:nussbaum_capabilities}). This information, corresponding to `context-specific data' in Fig.~\ref{fig:CAvalues}, could then be used as the baseline for modelling individual variations in the choice factor when initialising the agent population. 
%In order to emphasize the importance of the collective context in choice factors, we propose establishing a prioritisation of central capabilities to be restored (corresponding to the ``collective contextualized data" in Fig.~\ref{fig:CAvalues}). Their relative importance must be informed by the stakeholders (non-profits, PEH, or other community members), following a democratic and participatory process. Such collective information serves as the baseline for modelling individual variations in value preferences and need urgencies.
Overall, as can be seen in Fig.~\ref{fig:CAvalues}, our work considers that legal and social norms~\cite{Shoham1995} form the social conversion factors (from resources to capabilities). Also, we consider that context-specific data informs individual need urgencies and value preferences~\cite{schwartz}, which act as choice factors (from capabilities to functionings). %This integration aligns with the CA's emphasis on agency, choice, and freedom in evaluating social arrangements, policy impacts, and inequity. 

\textbf{Implementation.} From a computational perspective, we talk about possible, impossible and realised actions or goals, tied to short and long-term rewards. Possible (and impossible) actions represent capabilities (and deprived capabilities), whereas realised actions are the functionings in the CA literature. Short and long-term rewards represent the choice factors (need urgencies and value preferences) driving an agent's decision-making process. Actions seeking short-term rewards are driven by need urgencies, while actions seeking long-term rewards are driven by value preferences. The latter set of actions often enables the restoration of central capabilities (e.g. `securing a job' is an action seeking long-term rewards that enables \textit{control over one's environment} in Table~\ref{tab:nussbaum_capabilities}). Additionally, when we consider the dynamics of the decision-making, i.e., how the realised actions influence both the individual and overall system over time, two fundamental updating loops appear, represented in Fig.~\ref{fig:CAvalues}:

\begin{enumerate} [label=\arabic*)]
    \item \textbf{The impact of realised actions on resources and conversion factors}. Realised actions lead to updates on the agents' states. For instance, `riding a bike' might help me achieve the necessary conditions to `secure a job'. We build on~\cite{tseng2024ci} approach to take into consideration this updating between agents' behaviour and what they call technical and social context, which includes resources and conversion factors. %From a computational perspective, changes in functionings' prompt updates to resources and conversion factors, which include technical and social aspects influencing the well-being of individuals. The social context includes characteristics of individuals and communities that influence individuals’ ability to access physical and economic resources of a society. In \cite{tseng2024ci}, these are divided in human capital (knowledge, skills and health) and social capital (social relations, mutual trust, shared norms). The technical context includes several systems supporting community capitals transformation into goods and services essential to a community. In their framework, they overall consider the system of infrastructure systems but we can consider many other systems such as the healthcare system, educational system, social service system, industries system, etc.
\begin{comment}
\begin{figure*}[ht!]
    \centering
    \includegraphics[width=0.6\linewidth]{usecase (1).png}
    \caption{Representation of the Capability Approach operationalized for Health Inequity case study.}
    \label{fig:enter-label}
\end{figure*}
\end{comment}
%While choice factors inform which possible action (capability) an agent chooses to realise (functioning),
    \item \textbf{The impact of realised actions on choice factors}. Realised actions might also result in updating choice factors. For example, if I `have an accident while riding a bike' because I was careless, I might become more cautious next time. This implies that my preference for being healthy has increased, influencing the rewards functions of the decision-making. %While we acknowledge that choice factors evolve over time, to simplify the initial framework that we wish to implement, we leave this issue for future versions of the framework.  %THIS CHOICE FACTOR IS WITH THE STATE!!!! S' 
    %Over time, the value of security might become less prominent in her priorities, as the need is no longer a pressing concern.
\end{enumerate}

\subsection{The Capability Approach as a Markov Decision Process}

In reinforcement learning, the  Markov decision process (MDP) framework defines the interaction between a learning agent and its environment using states, actions, transition probabilities, and rewards. By mapping CA concepts in Fig.~\ref{fig:CAvalues} to MDP elements in Table~\ref{tab:my_label}, we enable the use of reinforcement learning to model the agents' decision-making. %While few scholars in the CA literature have used the approach in this way, it has been stated as a promising direction for future research that requires interdisciplinary effort.
%This capability theory aims to \emph{explain} poverty, inequality or wellbeing, helping us contextualize these notions, few scholars in the CA literature use the CA in this way, but it is known to be a strong potential future work there, that requires strongly discipliantory work. an explanatory purpose that has been largely ignored in the CA literature.  %under the constraints and multiobjective goals considered in the CA.

\begin{itemize}
    \item State $s$ represents the current situation the agent is in. In our framework, this includes almost all the information: the agent's personal characteristics, the available resources, and the individual need and value preferences.
    \item Actions are the set of possible decisions $a$ that the agent can make to interact with its environment. In our framework, we differentiate between possible, impossible and realised actions, analogous to specific capabilities, deprived capabilities and functionings. 
    \item Transition probabilities $P (s' | s, a)$ are the probabilities of moving from one state $s$ to another state $s'$ after taking action $a$. In our framework, transition probabilities may be used to represent conversion factors, or the probability of being capable of performing an action. One can imagine, for example, that impossible actions would have associated a 100\% chance of remaining in the same state, representing that nothing is happening. %moving from one state to another if an action was taken (not after an action is taken).
    \item Rewards are the feedback the agent receives when choosing action $a$ in a given state $s$. In our framework, this is how much have the choice factors (value preferences and need urgencies) been satisfied by that action. We differentiate between short-term and long-term sums of expected rewards, $Q_s(s, a')$ and $Q_l (s, a')$, to reflect urgency-based and importance-based outcomes. %In our framework, rewards are tied to individual choice factors, i.e. the agent earns higher rewards for actions assigned as most urgently needed or valued.
\end{itemize}
% \[
% Q(s, a) = \gamma^n \sum_{s'} P(s' \mid s, a_s) \max_{a'} Q(s', a') 
% \]

One main objective of our research is to study how these expected rewards are defined. For that, we need to analyse the dynamics between need urgencies and value preferences. The agent's policy $\pi (a | s)$ will define the probability of selecting action $a$ in state $s$ based on these two types of expected rewards   %, considering our two types of rewards
%The agent's policy $\pi (a | s)$ defines the probability of selecting action $a$ in state $s$ based on expected rewards:
% deterministic and greedy 
\[
\pi(a \mid s) =
\begin{cases}
%0, & \text{if } P(s \mid s, a) = 1 \text{ and } a \neq \text{do-nothing} \\
1, & \text{if } a = \arg\max_{a'} \oplus  (Q_s(s, a'), Q_l (s, a')) \\
0 & \text{otherwise},
\end{cases}
\]
where $\oplus$ indicates an aggregation of short-term and long-term reward components.

The novelty of the decision-making would precisely be defining this aggregation of needs and values for driving behaviour. Of course, different discount factors $\gamma \in [0,1]$ will need to be considered for the different reward types. This will allow to account for short-term and long-term effects of the agent's actions with lower or higher values of the discount factor, respectively. Depending on the results, we may also consider introducing constraints to specify that needs are addressed as long as they do not conflict with values. %All in all, the agent chooses the action that maximizes the expected reward, which depends on both need urgencies and value preferences.
%The MDP framework also includes transition probabilities (sometimes represented in transition matrices), which describe the likelihood of moving from one state to another after an action is taken in a given state. In particular, they are the probability that action a in state s at time t will lead to state s’ at time t’. If the number of actions is m, we will have m matrices. These transitions help the agent learn how its actions influence future states and rewards. In RL, the agent's goal is to learn a policy—a strategy for choosing actions in each state—to maximize its long-term rewards by exploring and exploiting its environment. 

 %Finally, rewards can be associated with the value promotion and need fulfilment of the choice factors. This reward defines a feedback loop between actions and the state after an agent has decided on an action and enjoys (or suffers) its effects.
\subsection{Towards Assessing Policies in terms of Capabilities with Agent-based Simulations}

The defined MDP will be the basis of the decision-making in the agent-based simulation. This simulation will allow to assess the impact of policies through the lens of capabilities. Building upon~\cite{aguilera2024can}, the simulation will be composed by (1) agents' profiles, (2) a physical environment and (3) a regulatory environment. By simulating agents' profiles, including personal conversion and choice factors instantiated from real-worl data, we will obtain a representation of the stakeholders involved, such as PEH, healthcare and social services, and non-profit organizations. The MDP will define their behaviour and social interactions based on these profiles. The physical and regulatory environment will determine their surroundings, including resources, social, and environmental conversion factors. In particular, legal norms will affect the agents' state and actions. The impact of such policies will be assessed by analyzing each individual's set of possible and impossible actions (enabling restored and deprived central capabilities) among other elements discussed in Section~\ref{sec:evaluation}.

\section{Foreseen Case Study}
Our proposed case study examines healthcare challenges in Barcelona's Raval neighbourhood, where the majority of PEH are located. According to \textit{Salut Sense Llar}, an organization of doctors specialized in treating PEH, the main problems they face include a systemic exclusion from primary healthcare (PHC), pharmaceutical poverty and a lack of post-discharge assistance~\cite{saumell2024atencion}. Despite suffering from lower life expectancy and higher morbidity~\cite{lahiguera2022analisis}, PEH encounter major barriers to accessing social and healthcare services. For instance, those in irregular administrative situations (non-registered citizens) are unable to access PHC. As a result, their health worsens over time, often reaching a critical point where they require emergency care and hospitalization.

Several studies show that integrating PHC in the management of PEH improved the diagnosis and treatment of chronic diseases while reducing visits to emergency services and hospital admissions~\cite{joyce2009,otoole2010,ponka2020}. Motivated by this evidence, the legal policies proposed by \textit{Salut Sense Llar} aim to address these health inequities, leading to personal suffering and both ethical and economic costs to society more broadly. Their goal is to offer improved healthcare to PEH by guaranteeing an inclusive PHC, with multidisciplinary teams, attention in situ, and a gender-sensitive approach. 

Roybens' modular view of the CA provides a robust lens to capture the healthcare challenges suffered by PEH. We start with a small-scale example, while acknowledging that its complexity should be enhanced for the use case to be truly informative. To begin with, we should consider multiple (if not all) central capabilities in Table~\ref{tab:nussbaum_capabilities} both in the rewards and the evaluative basis of the simulation. However, we begin by targeting \textit{bodily health} to show how the case study can later be developed within the project. 

%In this context, we illustrate 1) how the this case study can be implemented as an MDP, where agents' decision-making is driven by their needs and values, and 2) how simulating agents behaviour can provide insights into identifying policies that help restore central capabilities. %, in this case, focusing on one targeted central capability, bodily health in Table~\ref{tab:nussbaum_capabilities}. %this initial use case is merely illustrative, with very reduced state and action space, focusing primarily on bodily health. %In this example, we represent the most relevant elements to showcase the implementation of our framework. %As illustrated in Fig.~\ref{fig:enter-label}, we start by defining: However, we begin with a simplified representation that highlights the essential elements defined in our collaboration with non-profits in the healthcare domain.
\begin{figure}[ht]
    \centering
    \includegraphics[width=0.95\linewidth]{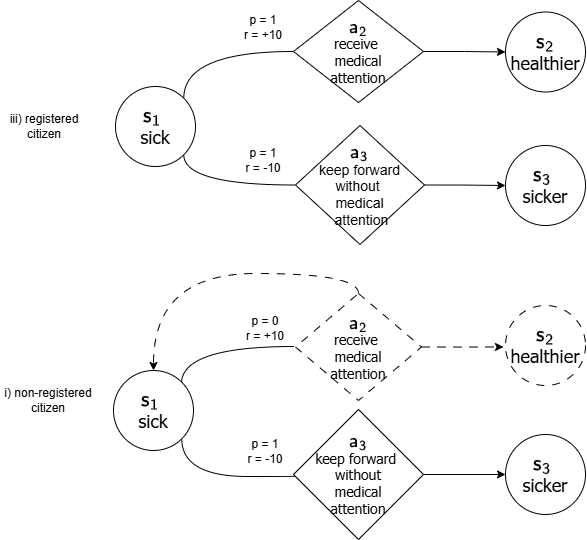}
    \caption{A Markov Decision Process (MDP) representation of the decision-making. Nodes represent states, while diamonds indicate actions that may be possible or impossible (i.e., capabilities or deprived capabilities) with lines and dashed lines, respectively.  } %Conversion factors are represented in the probabilities, while choice factors are represented in the rewards.
    \label{fig:MDP}
\end{figure}
\subsection{Health Inequity as a CA-based MDP}
\label{sec: mdpusecase}
We focus on two scenarios: a sick registered citizen and a sick non-registered citizen. The difference lies in how a legal norm constrains their access to primary healthcare (PHC). Figure~\ref{fig:MDP} illustrates how the situation can be modeled as an MDP. We make three main assumptions: %concerning the action and state space, the transition probabilities and the rewards: 

\begin{enumerate}[label=\arabic*)]
    \item We assume a reduced action and state space: the only relevant information in the agents' personal profile is their health and registration state, and the only executable actions are `receive medical attention' and `keep forward without medical attention'.
    \item We assume that transition probabilities are binary: receiving medical attention unequivocally improves the agent's health state, and the legal norm unequivocally determines whether the agent receives medical attention or not.
    \item We assume simplified reward functions: we consider the agent highly values \emph{bodily health}, enabled by the action `receive medical attention'. We do not consider urgency-based actions or other central capabilities that could be enabled by the action `keep forward without medical attention'. %In this scenario, we only consider the influence of values, as the actions in question are importance-based actions tied to long-term rewards.
\end{enumerate}
And we consider the following elements:
\begin{enumerate}
    \item Resources: primary healthcare (PHC) services.
    \item Conversion factors / Transition Probabilities defining the possibility of actions: 
    \begin{enumerate}
        \item Personal: health and registration state.
        \item Social: legal norm specifying who gets access to PHC based on registration state.
    \end{enumerate}
    \item Capabilities: 
    \begin{enumerate}

        \item Specific capabilities / Possible actions: being able to receive medical attention or not.
        \item Central capabilities to be restored / Long-term goals: Bodily health.
    \end{enumerate}
    \item Choice factors / Rewards dictating decision-making:
    \begin{enumerate}
        \item Value preferences: `I value being healthy'.
        \item Need urgencies: `I am in pain'.
    \end{enumerate}
    \item Functioning / Realised action: receive medical attention or not.
\end{enumerate}

Under these assumptions, the MDP starts with the agent in a sick state, $s_1$. For a registered citizen, two actions are possible: (i) $a_2$ `receive medical attention' leads to a healthier state $s_2$ with reward $r= +10$, and (ii) $a_3$ `keep forward without medical attention' leads to a sicker state $s_3$ with reward $r=-10$. Both transitions occur with probabilities $p=1$, meaning that the actions will unequivocally lead the agent to the associated state. Because the agent highly values \emph{bodily health}, action $a_2$ is prioritised over action $a_3$ and the optimal policy of the agent will make him end up in a healthier state $s_2$. For a non-registered citizen, $a_2$ is an impossible action (deprived capability), although it is a highly valued goal with reward $r= +10$. The only possible action is $a_3$ which leads to a sicker state $s_3$ with negative reward $r=-10$. 
%However, imagine that the agent is being threatened to not go to the hospital, and he values bodily integrity above bodily health.?? This means another action  above this one

\subsection{Practical Implementation}
\label{subsec: futimpl}
%By enhancing the complexity of such scenarios, we could quantify these disparities and explore alternative policies that restore deprived capabilities, enabling individuals to further expand their capabilities.
The presented MDP demonstrates how can we model behaviour being affected by policies that lead to inequitable health outcomes. This project aims to define in detail the rest of the relevant elements listed in Section~\ref{sec: mdpusecase} by working closely with the stakeholders involved in this context. For the practical implementation, we will simulate agent profiles by relying on existing anonymised data collected annually by non-profits~\cite{ArrelsRecompte2023}. By applying a probabilistic approach similar to~\cite{aguilera2024value}, we will sample a synthetic population from this data. Additionally, choice factors will be informed by context-specific data to guide the prioritisation of central capabilities. For instance, we will use prioritisations based on needs expressed by PEH in~\cite{WorldBankVoices2000}, and interviews from~\cite{caritas2021informe}. These sources indicate that \emph{bodily integrity} and \emph{bodily health} hold the same priority level as \emph{affiliation} or \emph{control over one's environment}. On the other hand, \emph{life}, \emph{senses, imagination and thought} or \emph{play} are considered at a secondary level. Such information can help us tune the individual variations in the choice factor, ensuring the agent's behaviour aligns with lived experiencies and priorities of PEH.

\section{Implementation Plans, Evaluation and Expected Social Impact}
\label{sec:evaluation}
Our implementation plan begins with the development of the MDP for the health inequity use case. We will start small with the example, and then build the complexity of the model as we progress. This will give us time to address challenges in obtaining all the data we need (for example, the necessary data and procedures to obtain representative profiles of PEH, mentioned in subsection~\ref{subsec: futimpl}), as well as challenges in modelling (for example, defining the interactions between needs and values within the decision-making process of the agent). All this work will be carried out in close collaboration with selected nonprofit organizations and other domain experts in human development studies and PEH's healthcare. It will be of utmost importance to present to them our model's mechanism and results (in focus groups), to collect feedback that guides our work on enhancing the model's complexity.

As outlined by the CA, capabilities and functionings are essential but not exclusive elements of evaluation. To assess the simulation outcomes, we consider other indicators that help us reflect the multidimensions of homelessness: (1) the housing state (following ETHOS~\cite{amore2011ethos} terminology), (2) the health state (establishing a quantitative scale with the pathologies described in~\cite{saumell2024atencion}), and (3) the registration state (registered, non-registered or in process, following the descriptions of social services workers). Additionally, we will evaluate (4) governmental economic expenses for both healthcare services and social services, as well as (5) discrimination within institutional frameworks, together with values being promoted or demoted by legal and social norms. 

Most of the above can be evaluated by checking the state in the MDP, such as the housing or health state of an agent, or the expense state of different organisations. In order to evaluate capabilities and functionings, we will check the action space: possible, impossible and realised actions enabling central capabilities in Table.~\ref{tab:nussbaum_capabilities}. %Discrimination will be the tricky indicator, as we first need to carefully define it computationally in order to evaluate it. 

The evaluation criteria defined aligns with the multidimensionality of the CA. It is also aligned with the LNOB principle, which underscores that discrimination and inequalities (often multiple and intersecting) undermine the agency of people as holders of rights~\cite{UN2023LeaveNoOneBehind}. The expected simulation results would impact multiple SDGs: No Poverty (1), Zero Hunger (2), Good Health and Well-Being (3), Decent Work and Economic Growth (8) Industry Innovation and Infrastructure (9), Sustainable Cities and Communities (11), and Partnerships (17)~\cite{sdgs2023}.

\section{Challenges, Limitations and Ethical Considerations}
As demonstrated through our work, the CA can be used in many different directions but needs additional specifications to become effective~\cite{robyens2017}. This makes it particularly suitable for computational modelling. However, it heavily relies on interdisciplinary collaboration to define the context-specific elements of the CA depending on the case study. For that, we emphasize the importance of following Roybens' modular view of the CA~\cite{robyens2017}. These modules include mandatory specification of key elements in the CA (conversion factors, capabilities and functionings, other dimensions of evaluative value, etc.), as well as accounts for human diversity, agency, and structural constraints. Given the need for context-specific detail, our framework will be designed to be flexible yet general enough to be applied in diverse settings where inequities arise.

As we enhance the complexity of our model, including larger state and action spaces or bigger population, scalability becomes a significant challenge. A simple MDP approach can struggle with exponential growth in states and actions, especially when considering impossible actions too (i.e., deprived capabilities). To address these computational limitations, our work will likely employ function approximation or hierarchical methods, combined with parallelization techniques on the Ars manga cluster~\cite{iiia} to maintain economic and energetic sustainability.

From an ethical standpoint, we plan to rely on anonymized and synthetic data, ensuring privacy and confidentiality. Additionally, we emphasize the relevance of discrimination in our framework. In line with the principle of LNOB, we acknowledge that many barriers to accessing services, resources and equal opportunities are not simply a lack of availability of resources, but rather the result of discriminatory laws, policies and social practices that leave particular groups of people further and further behind~\cite{UN2023LeaveNoOneBehind}. Our entire proposal is focused on addressing these challenges. We will pay particular attention to aporophobia (fear or rejection towards the poor)~\cite{cortina2017aporofobia} and its impact throughout our research. If needed, we also plan to seek the approval from CSIC's ethics committee. 

% \begin{itemize}
%     \item Contextualizaiton of everything is mandatory! 
%     \item Limitations: Computational space limitations. Additional specifications ......
%     \item Scalability: This framework is designed to be adapted to any context where inequity arises. However, the computational limitations can affect the amount of population one wants to consider.
%     \item Economic sustainability of the solution: Parallelization to save up energy int he cluster?
%     \item Ethical considerations: privacy and confidentiality, discrimination, such as aporophobia, should be central!!! State whether you will seek (or have already obtained) approval from an Institutional Review Board (IRB), Ethics Committee, or comparable body.
% \end{itemize}

% Tehnical limitations: coherently integrate values with needs in the simulation. coherent dynamic of the expansion of capabilities. computational limitations in ABM

%In the ABM and game theory domain, agency has a different connotation: it is the ability of an agent to act independently and make choices to achieve their goals or maximize their utility. Therefore, agency from the CA perspective is more complex in the sense that considers that maximizing your utility sometimes is dependent on maximizing other utilities (even if it means minimizing yours). That's where values can enter the picture, too. 

\appendix

\section*{Acknowledgments}

This research has been supported by the EU-funded VALAWAI (\#~101070930), the Spanish-funded VAE (\#~TED2021-131295B-C31) and the Rhymas (\#~PID2020-113594RB-100) projects. Special thanks to all the local stakeholders involved. Beatriz Férnandez, from Fundació Arrels, for sharing her law proposal, and Beatriu Bilbeny, from Salutsensellar, for guiding us toward identifying the key issues to address. Thanks to the human development community, including Flavio Comin and Mark Fabian, for giving us the necessary feedback to carry on with the proposal.  %We also appreciate the contributions of the social services workers, Nuri Ferran and Bet [last name].

\bibliographystyle{named}
\bibliography{ijcai24}

\begin{thebibliography}{}

\bibitem[\protect\citeauthoryear{Acemoglu and Robinson}{2012}]{acemoglu2012}
Daron Acemoglu and James~A. Robinson.
\newblock {\em {Why Nations Fail: The Origins of Power, Prosperity, and Poverty}}.
\newblock {Crown Business}, {New York, NY, USA}, 2012.

\bibitem[\protect\citeauthoryear{Aguilera \bgroup \em et al.\egroup }{2024}]{aguilera2024can}
Alba Aguilera, Nieves Montes, Georgina Curto, Carles Sierra, and Nardine Osman.
\newblock Can poverty be reduced by acting on discrimination? an agent-based model for policy making.
\newblock {\em arXiv preprint arXiv:2403.01600}, 2024.

\bibitem[\protect\citeauthoryear{Aguilera \bgroup \em et al.\egroup }{2025}]{aguilera2024value}
Alba Aguilera, Miquel Albert{\'\i}, Nardine Osman, and Georgina Curto.
\newblock Value-enriched population synthesis: Integrating a motivational layer.
\newblock In Nardine Osman and Luc Steels, editors, {\em Value Engineering in Artificial Intelligence - Second International Workshop, {VALE} 2024, Santiago de Compostela, Spain, October 19, 2024, Proceedings}, volume 15356 of {\em Lecture Notes in Computer Science}, pages 1--20. Springer, 2025.

\bibitem[\protect\citeauthoryear{Allport}{1954}]{allport1954}
Gordon~W. Allport.
\newblock {\em {The Nature of Prejudice}}.
\newblock {Addison-Wesley}, {Reading, MA, USA}, 1954.

\bibitem[\protect\citeauthoryear{Amore \bgroup \em et al.\egroup }{2011}]{amore2011ethos}
Kate Amore, Michael Baker, and Philippa Howden-Chapman.
\newblock The ethos definition and classification of homelessness: An analysis.
\newblock {\em European Journal of Homelessness}, 5(2):19--37, 2011.

\bibitem[\protect\citeauthoryear{{Arrels Fundació}}{2023}]{ArrelsRecompte2023}
{Arrels Fundació}.
\newblock Recompte 2023: Resultats, 2023.
\newblock Consultat el 10 de febrer de 2025.

\bibitem[\protect\citeauthoryear{{Artificial Intelligence Research Institute (IIIA-CSIC)}}{2025}]{iiia}
{Artificial Intelligence Research Institute (IIIA-CSIC)}.
\newblock Servicios de computación.
\newblock \url{https://www.iiia.csic.es/es/research/servicios-de-computacion/}, 2025.

\bibitem[\protect\citeauthoryear{Assa and Lengfelder}{2020}]{assa2020can}
Jacob Assa and Christina Lengfelder.
\newblock Can enhancing capabilities promote energy justice? an agent-based model approach.
\newblock {\em Mendeley Data}, 1, 2020.

\bibitem[\protect\citeauthoryear{Ch{\'a}vez-Ju{\'a}rez and Krishnakumar}{2021}]{chavez2021capmod}
Florian Ch{\'a}vez-Ju{\'a}rez and Jaya Krishnakumar.
\newblock Capmod: a simulated society to evaluate empirical estimators of capabilities.
\newblock {\em Journal of Human Development and Capabilities}, 22(1):52--79, 2021.

\bibitem[\protect\citeauthoryear{{City and County of San Francisco}}{2024}]{sf_homeless_population_2024}
{City and County of San Francisco}.
\newblock {Homeless Population}.
\newblock {San Francisco Government Website}, 2024.
\newblock [Accessed 22 Jan 2025].

\bibitem[\protect\citeauthoryear{Cortina}{2017}]{cortina2017aporofobia}
Adela Cortina.
\newblock {\em Aporofobia, el rechazo al pobre}.
\newblock Paidós, Barcelona, Spain, 2017.

\bibitem[\protect\citeauthoryear{{Cáritas}}{2021}]{caritas2021informe}
{Cáritas}.
\newblock {ANEXO 1 Informe 2021 -- Relatorías Especiales de Naciones Unidas sobre la Extrema Pobreza y los Derechos Humanos y sobre el Derecho a una Vivienda Adecuada}, 2021.
\newblock Publicación de Cáritas.

\bibitem[\protect\citeauthoryear{De~Wildt \bgroup \em et al.\egroup }{2020}]{de2020conflicted}
TE~De~Wildt, EJL Chappin, G~van~de Kaa, PM~Herder, and IR~van~de Poel.
\newblock Conflicted by decarbonisation: Five types of conflict at the nexus of capabilities and decentralised energy systems identified with an agent-based model.
\newblock {\em Energy Research \& Social Science}, 64:101451, 2020.

\bibitem[\protect\citeauthoryear{Dignum}{2021}]{dignum2021social}
Frank Dignum.
\newblock {\em Social simulation for a crisis}.
\newblock Springer, 2021.

\bibitem[\protect\citeauthoryear{{European Commission}}{2021}]{EuropeanCommission_Homelessness}
{European Commission}.
\newblock Homelessness.
\newblock \url{https://employment-social-affairs.ec.europa.eu/policies-and-activities/social-protection-social-inclusion/addressing-poverty-and-supporting-social-inclusion/homelessness_en}, 2021.
\newblock Accessed: 2025-02-12.

\bibitem[\protect\citeauthoryear{{FEANTSA and Abbé Pierre Foundation}}{2023}]{feantsa2023overview}
{FEANTSA and Abbé Pierre Foundation}.
\newblock {8th Overview of Housing Exclusion in Europe 2023}.
\newblock {FEANTSA, Brussels}, 2023.
\newblock {[Accessed 21 Jan 2025]}.

\bibitem[\protect\citeauthoryear{Gasper}{2017}]{gasper2017sdgs}
Des Gasper.
\newblock The {SDGs} and the capability approach: Concerns and opportunities.
\newblock {\em Journal of Human Development and Capabilities}, 18(3):355--361, 2017.

\bibitem[\protect\citeauthoryear{Honneth}{1996}]{honneth1996}
Axel Honneth.
\newblock {\em {The Struggle for Recognition: The Moral Grammar of Social Conflicts}}.
\newblock {MIT Press}, {Cambridge, MA, USA}, 1996.

\bibitem[\protect\citeauthoryear{Joyce and Limbos}{2009}]{joyce2009}
D.P. Joyce and M.~Limbos.
\newblock {Identification of cognitive impairment and mental illness in elderly homeless men: Before and after access to primary health care}.
\newblock {\em Canadian Family Physician}, 55:1110--1111, 2009.

\bibitem[\protect\citeauthoryear{Lahiguera \bgroup \em et al.\egroup }{2022}]{lahiguera2022analisis}
Daniel~Roca Lahiguera, Beatriu~Bilbeny de~Fortuny, del CAP Raval~Sud Gironella~et.al., Grupo de~Estudio~del Sinhogarismo, et~al.
\newblock An{\'a}lisis de la salud de la poblaci{\'o}n sin hogar de un distrito desfavorecido de barcelona. estudio essella.
\newblock {\em Atenci{\'o}n primaria}, 54(10):102458, 2022.

\bibitem[\protect\citeauthoryear{Markhvida \bgroup \em et al.\egroup }{2020}]{markhvida2020quantification}
Maryia Markhvida, Brian Walsh, Stephane Hallegatte, and Jack Baker.
\newblock Quantification of disaster impacts through household well-being losses.
\newblock {\em Nature Sustainability}, 3(7):538--547, 2020.

\bibitem[\protect\citeauthoryear{Marshall}{2024}]{marshall2024beyond}
Carrie Anne et~al. Marshall.
\newblock Beyond securing a tenancy: Using the capabilities approach to identify the daily living needs of individuals during and following homelessness.
\newblock {\em Journal of Social Distress and Homelessness}, 33(1):81--95, 2024.

\bibitem[\protect\citeauthoryear{Maslow}{1943}]{maslow1943needs}
Abraham~H. Maslow.
\newblock {\em A Theory of Human Motivation}, volume~50.
\newblock Psychological Review, 1943.

\bibitem[\protect\citeauthoryear{Melin \bgroup \em et al.\egroup }{2021}]{melin2021energy}
Anders Melin, Rosie Day, and Kirsten~EH Jenkins.
\newblock Energy justice and the capability approach—introduction to the special issue.
\newblock {\em Journal of Human Development and Capabilities}, 22(2):185--196, 2021.

\bibitem[\protect\citeauthoryear{Nussbaum}{2011}]{nussbaum2011}
Martha~C. Nussbaum.
\newblock {\em {Creating Capabilities: The Human Development Approach}}.
\newblock {Harvard University Press}, {Cambridge, MA, USA}, 2011.

\bibitem[\protect\citeauthoryear{{OECD}}{2024}]{oecd2024homelessness}
{OECD}.
\newblock {OECD Toolkit to Combat Homelessness}.
\newblock {OECD, Paris}, 2024.
\newblock {[Accessed 21 Jan 2025]}.

\bibitem[\protect\citeauthoryear{Osman and d'Inverno}{2024}]{nardine}
Nardine Osman and Mark d'Inverno.
\newblock A computational framework of human values.
\newblock In Mehdi Dastani, Jaime~Sim{\~{a}}o Sichman, Natasha Alechina, and Virginia Dignum, editors, {\em Proceedings of the 23rd International Conference on Autonomous Agents and Multiagent Systems, {AAMAS} 2024, Auckland, New Zealand, May 6-10, 2024}, pages 1531--1539. International Foundation for Autonomous Agents and Multiagent Systems / {ACM}, 2024.

\bibitem[\protect\citeauthoryear{O'Toole}{2010}]{otoole2010}
et~al. O'Toole, T.P.
\newblock {Applying the chronic care model to homeless veterans: Effect of a population approach to primary care on utilization and clinical outcomes}.
\newblock {\em American Journal of Public Health}, 100:2493--2499, 2010.

\bibitem[\protect\citeauthoryear{Piketty}{2017}]{piketty2017}
Thomas Piketty.
\newblock {\em {Capital in the Twenty-First Century}}.
\newblock {Harvard University Press}, {Cambridge, MA, USA}, 2017.

\bibitem[\protect\citeauthoryear{Ponka and Agbata}{2020}]{ponka2020}
D.~Ponka and E.~et~al. Agbata.
\newblock {The effectiveness of case management interventions for the homeless, vulnerably housed and persons with lived experience: a systematic review}.
\newblock {\em PLoS ONE}, 15(4):e0230896, 2020.

\bibitem[\protect\citeauthoryear{Robeyns}{2005}]{robeyns2005capability}
Ingrid Robeyns.
\newblock The capability approach: a theoretical survey.
\newblock {\em Journal of Human Development}, 6(1):93--117, 2005.

\bibitem[\protect\citeauthoryear{Robeyns}{2017}]{robyens2017}
Ingrid Robeyns.
\newblock {\em {Wellbeing, Freedom and Social Justice: The Capability Approach Re-Examined}}.
\newblock {Open Book Publishers}, {Cambridge, UK}, 2017.

\bibitem[\protect\citeauthoryear{Sandel}{2020}]{sandel2020}
Michael~J. Sandel.
\newblock {\em {The Tyranny of Merit: What's Become of the Common Good?}}
\newblock {Farrar, Straus and Giroux}, {New York, NY, USA}, 2020.

\bibitem[\protect\citeauthoryear{{Sant Joan de Déu Serveis Socials Barcelona}}{2023}]{sjd2023}
{Sant Joan de Déu Serveis Socials Barcelona}.
\newblock {Resum de la llei 2023}.
\newblock {Sant Joan de Déu Serveis Socials Barcelona}, 2023.
\newblock {[Accessed 21 Jan 2025]}.

\bibitem[\protect\citeauthoryear{Saumell \bgroup \em et al.\egroup }{2024}]{saumell2024atencion}
Carme~Roca Saumell, Sergio~Moreno Ferrer, Mar{\'\i}a-Paz~Loscertales de~la Puebla, Beatriu~Bilbeny de~Fortuny, and Jordi~Del{\'a}s Amat.
\newblock Atenci{\'o}n sanitaria a las personas sin hogar.
\newblock {\em FMC-Formaci{\'o}n M{\'e}dica Continuada en Atenci{\'o}n Primaria}, 31(3):118--123, 2024.

\bibitem[\protect\citeauthoryear{Schwartz}{1992}]{schwartz}
Shalom~H. Schwartz.
\newblock Universals in the content and structure of values: Theoretical advances and empirical tests in 20 countries.
\newblock {\em Advances in Experimental Social Psychology}, 25:1--65, 1992.

\bibitem[\protect\citeauthoryear{Schwartz}{2012}]{schwartz2012overview}
SH~Schwartz.
\newblock An overview of the schwar view of the schwartz theor tz theory of basic vy of basic values.
\newblock {\em Online readings in Psychology and Culture}, 2(1):11--20, 2012.

\bibitem[\protect\citeauthoryear{Sen}{1979}]{sen1979}
Amartya Sen.
\newblock {Equality of What?}
\newblock In Sterling~M. McMurrin, editor, {\em {The Tanner Lectures on Human Values}}, pages 195--220. {University of Utah Press}, {Salt Lake City, UT, USA}, 1979.

\bibitem[\protect\citeauthoryear{Sen}{1999}]{sen1999}
Amartya Sen.
\newblock {\em {Development as Freedom}}.
\newblock {Oxford University Press}, {Oxford, UK}, 1999.

\bibitem[\protect\citeauthoryear{Shoham and Tennenholtz}{1995}]{Shoham1995}
Yoav Shoham and Moshe Tennenholtz.
\newblock On social laws for artificial agent societies: off-line design.
\newblock {\em Artificial Intelligence}, 73(1-2):231--252, 1995.

\bibitem[\protect\citeauthoryear{Silva-Lopez \bgroup \em et al.\egroup }{2022}]{silva2022commuter}
Rodrigo Silva-Lopez, Gitanjali Bhattacharjee, Alan Poulos, and Jack~W Baker.
\newblock Commuter welfare-based probabilistic seismic risk assessment of regional road networks.
\newblock {\em Reliability Engineering \& System Safety}, 227:108730, 2022.

\bibitem[\protect\citeauthoryear{Taylor}{1989}]{taylor1989}
Charles Taylor.
\newblock {\em {The Sources of the Self: The Making of Modern Identity}}.
\newblock {Harvard University Press}, {Cambridge, MA, USA}, 1989.

\bibitem[\protect\citeauthoryear{{Trust for London}}{2023}]{trustforlondon2023}
{Trust for London}.
\newblock {Rough Sleeping in London: Data and Trends}.
\newblock {Trust for London}, 2023.
\newblock {[Accessed 21 Jan 2025]}.

\bibitem[\protect\citeauthoryear{Tseng and Stojadinovi{\'c}}{2024}]{tseng2024ci}
Ting-Hsiang Tseng and Bo{\v{z}}idar Stojadinovi{\'c}.
\newblock Ci-str: A capabilities-based interface to model socio-technical systems in disaster resilience assessment.
\newblock {\em International Journal of Disaster Risk Reduction}, 111:104763, 2024.

\bibitem[\protect\citeauthoryear{{United Nations Sustainable Development Group (UNSDG)}}{2023}]{UN2023LeaveNoOneBehind}
{United Nations Sustainable Development Group (UNSDG)}.
\newblock Leave no one behind.
\newblock \url{https://unsdg.un.org/2030-agenda/universal-values/leave-no-one-behind}, 2023.
\newblock Accessed: September 12, 2023.

\bibitem[\protect\citeauthoryear{{United Nations}}{1948}]{un1948udhr}
{United Nations}.
\newblock {Universal Declaration of Human Rights}.
\newblock General Assembly resolution 217 A, adopted on 10 December 1948, 1948.
\newblock Accessed: [insert date of access].

\bibitem[\protect\citeauthoryear{{United Nations}}{2023}]{sdgs2023}
{United Nations}.
\newblock {Accelerating the 2030 Agenda for Sustainable Development: Center for Comprehensive Homeless Services}.
\newblock {Sustainable Development Goals Partnership Platform}, 2023.
\newblock {[Accessed 21 Jan 2025]}.

\bibitem[\protect\citeauthoryear{Williams \bgroup \em et al.\egroup }{2022}]{williams2022integrating}
Tim~G Williams, Daniel~G Brown, Seth~D Guikema, Tom~M Logan, Nicholas~R Magliocca, Birgit M{\"u}ller, and Cara~E Steger.
\newblock Integrating equity considerations into agent-based modeling: A conceptual framework and practical guidance.
\newblock {\em Journal of Artificial Societies and Social Simulation}, 25(3), 2022.

\bibitem[\protect\citeauthoryear{{World Bank}}{2000}]{WorldBankVoices2000}
{World Bank}.
\newblock {\em Voices of the Poor}.
\newblock World Bank Publications, Washington, DC, 2000.
\newblock \url{https://openknowledge.worldbank.org/handle/10986/13850}.

\bibitem[\protect\citeauthoryear{{Xarxa d'Atenció a Persones Sense Llar (XAPSLL)}}{2023}]{diagnosi2022barcelona}
{Xarxa d'Atenció a Persones Sense Llar (XAPSLL)}.
\newblock {Diagnosis 2022: Homelessness in Barcelona}.
\newblock {Barcelona Support Network for the Homeless (XAPSLL), Barcelona}, 2023.
\newblock {[Accessed 21 Jan 2025]}.

\end{thebibliography}

\clearpage  % force a new page for the CV section

%------------------------------------
% About the Authors / Team CV Section
%------------------------------------

\end{document}